\documentclass[acmsmall]{acmart}
\usepackage{paper_alcides}
\usepackage[capitalize,noabbrev]{cleveref}
\usepackage{graphicx}
\usepackage{enumitem}
\usepackage{caption}
\usepackage{subcaption}
\usepackage{csquotes}

\AtBeginDocument{%
  \providecommand\BibTeX{{%
    \normalfont B\kern-0.5em{\scshape i\kern-0.25em b}\kern-0.8em\TeX}}}

\begin{document}


\title{User-driven Design and Evaluation of Liquid Types in Java}



%
\author{Catarina Gamboa}
\email{cvgamboa@fc.ul.pt}
\affiliation{
  \institution{LASIGE, Faculdade de Ci\^{e}ncias da Universidade de Lisboa}
  \city{Lisboa}
  \country{Portugal}
}

\author{Paulo Alexandre Santos}
\email{pacsantos@fc.ul.pt}
\affiliation{
  \institution{LASIGE, Faculdade de Ci\^{e}ncias da Universidade de Lisboa}
  \city{Lisboa}
  \country{Portugal}
}

\author{Christopher S. Timperley}
\email{ctimperley@cmu.edu}
\affiliation{
  \institution{School of Computer Science, Carnegie Mellon University}
  \city{Pittsburgh}
  \country{USA}
}

\author{Alcides Fonseca}
\email{amfonseca@fc.ul.pt}
\affiliation{
  \institution{LASIGE, Faculdade de Ci\^{e}ncias da Universidade de Lisboa}
  \city{Lisboa}
  \country{Portugal}
}

\begin{abstract}

Bugs that are detected earlier during the development lifecycle are easier and cheaper to fix, whereas
bugs that are found during production are difficult and expensive to address, and may have dire consequences.
Type systems are particularly effective at identifying and preventing bugs early in the development lifecycle by
causing invalid programs to result in build failure.

Liquid Types are more powerful than those found in mainstream programming languages, allowing the detection of more classes of bugs.
However, while Liquid Types were proposed in 2008 with their integration in ML and subsequently introduced in C (2012), Javascript(2012) and Haskell(2014) through language extensions, they have yet to become widely adopted by mainstream developers.
This paper investigates how Liquid Types can be integrated in a mainstream programming language, Java, by proposing a new design that aims to lower the barrier to entry and adapts to problems that Java developers commonly encounter at runtime.

To promote accessibility, we conducted a series of developer surveys to design the syntax of LiquidJava, our prototype.
To evaluate the prototype's usability, we conducted a user study of 30 Java developers, concluding that users intend to use LiquidJava and that it helped to find more bugs and debug faster.
\end{abstract}



\maketitle

\section{Introduction}
Software quality is a major concern throughout software development~\cite{roleVV}.
Given the increased costs of finding and addressing bugs later in the development lifecycle,
many developers and organizations aim to identify issues as early as possible when they are cheaper
and easier to address by \enquote{shifting left}~\cite{SEGoogle}.
Unsurprisingly, validation techniques that are integrated into editors have proved widely popular
and been adopted by many developers.

Strong type systems have been introduced in modern programming languages (e.g., Haskell, Java), allowing developers to specify the expected type of operations and verify, at compile time, if those types are respected.
Editors typically integrate this verification to
provide developers with immediate feedback about the incorrect use of variables and values, allowing them to
debug and resolve the cause of identified issues more quickly.

Refinement types are more expressive than the relatively simple types of these programming languages~\cite{jhala2020refinement},
as they extend the language with predicates that restrict the allowed values in variables and methods.
For example, \cref{code:simple} shows a simple refinement on the variable \lstinline|r,| stating that its value must be between 0 and 255.
Refinement types can be used to detect simple division by zero errors and out-of-bounds array access bugs~\cite{xi1998arraybounds},
but they have also been used to detect security issues~\cite{bengtson2008refinements} and protocol violations~\cite{DBLP:conf/sefm/BurnayLV20}.

\begin{lstlisting}[caption=Variable refinement in LiquidJava and verification of its assignments., label=code:simple, basicstyle=\small]
@Refinement("r >= 0 && r <= 255")
int r;
r = 90; // okay
r = 200 + 60; /* okay in Java, Refinement Type Error: 
            Type expected: (r >= 0 && r <= 255); 
            Refinement found: (r == 200 + 60)*/
\end{lstlisting}

Although refinement types have been introduced in programming languages such as ML~\cite{refinements-ml}, Haskell~\cite{vazou2014refinement}, C~\cite{csolve} and Javascript\cite{DJS},
they have not yet been widely adopted by developers.
There are many plausible explanations for the lack of adoption, including, but not limited to,
a lack of awareness about refinement types and the language extensions that provide them,
developers may lack experience in writing specifications,
or it could be that the first languages to support refinement types were not widely used within
the industry at the time.

The primary goal of this work is to explore how to promote the wide usage of
refinement types by adding them to Java, one of the most popular programming
languages in the world, and using developer feedback to improve the usability
of refinement types in general.


\section{LiquidJava Design}
LiquidJava represents the integration of Liquid Types in Java, having a design focused on usability through the identification of system requirements (\cref{subsec:requirements}) and the guidance of the refinements language by Java developers (\cref{subsec:syntax-survey}). The language features are presented in~\cref{subsec:liquidjava_features} and aim to adapt to problems that Java developers commonly encounter at runtime. Finally, we implemented a prototype for LiquidJava and integrated it into an IDE to improve the usability of the framework (\cref{subsec:ide}).

\subsection{Requirements}\label{subsec:requirements}
To promote the usability of refinement types, we identified the following design requirements based on popular characteristics of successful static verification techniques (e.g., the NonNull annotation~\cite{DBLP:conf/oopsla/FahndrichL03}) and feedback from developers~\cite{DBLP:journals/cacm/SadowskiAEMJ18}:
\begin{itemize}
  \item \textbf{Refinements are optional.} 
  A valid Java program without refinements should type-check,\label{requirement:optional} allowing developers to incrementally adopt stronger types. This requirement can be fulfilled by using parametric annotations, such as the \lstinline|@Refinement| annotation;
  \item \textbf{Idiomatic refinements.} 
  The language of the refinements should be as close to Java as possible, so the developer can intuitively use and write the specification without learning a different language. To gain flexibility, we encoded the refinements language as strings inside the annotations. We created an online survey, detailed in \cref{subsec:syntax-survey}, to access the best syntax for the refinements, and developed the grammar based on the survey results;\label{requirement:expressive}
  \item \textbf{Refinement type-checking should be decidable} 
  without introducing unnecessary overhead to the compilation process and providing interactive feedback to developers as they type. To this end, we restricted the refinements language to Liquid Types~\cite{liquidtypes} which are verifiable by SMT solvers;\label{requirement:decidable} 
  \item \textbf{Refinement type-checking works on top of an existing Java type-checker} 
  to avoid early deprecation as the Java release cycle increases.
  Therefore, we implemented the refinement type checker on top of Spoon~\cite{DBLP:journals/spe/PawlakMPNS16} which uses the Eclipse Compiler Kit, adopting Java features as they are released.\label{requirement:ontop-java}
\end{itemize}

These requirements guided the design of LiquidJava and the implementation of a
prototype of the refinement type checker that was then integrated as an IDE
plugin.

\subsection{Syntax Survey}\label{subsec:syntax-survey}
\begin{figure}[htb]
  \centering
  \includegraphics[scale=0.35]{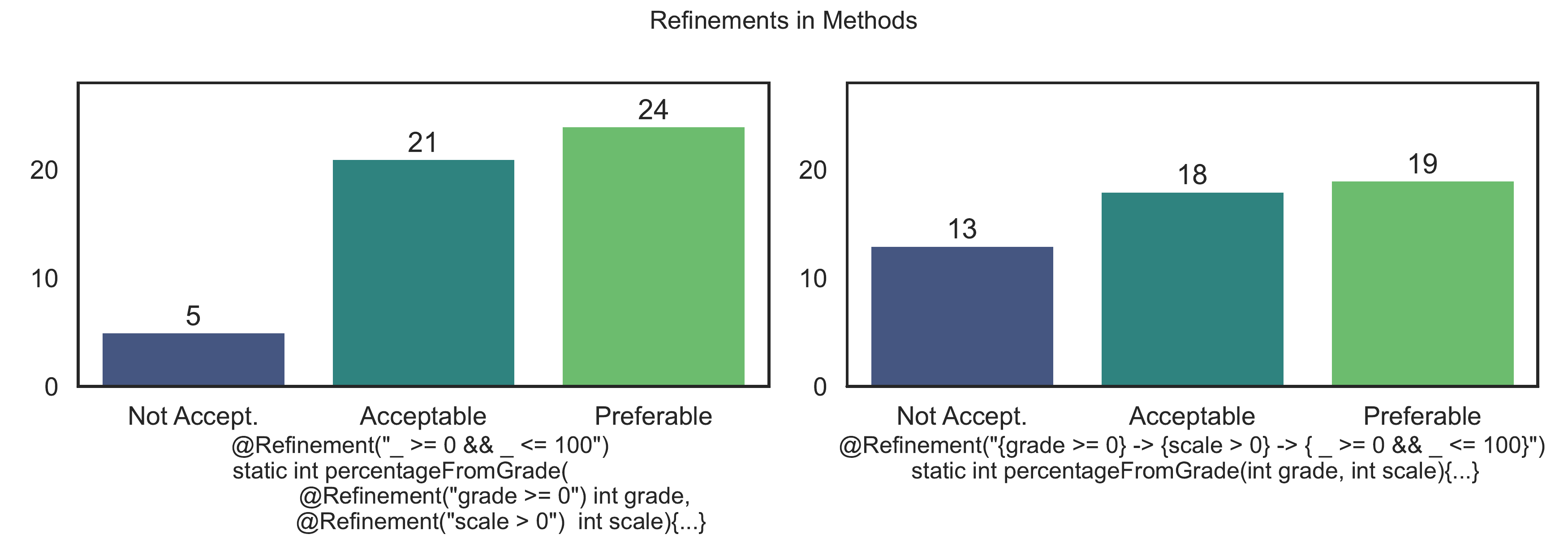}
  \caption{Preferences on the Syntax for Methods Refinements.}
  \label{fig:syntax-survey-answer-method}
\end{figure}

To assess the best syntax of the refinements language, we created and shared an online survey with multiple possible syntaxes for different LiquidJava features: type refinements of variables and methods, predicate aliases, and anonymous variables. The survey had 50 valid answers from participants familiar with Java, from which the majority was "Not Familiar" with refinement types.
Participants evaluated their preference for the syntax options, and the options that were most preferable were used to guide the development of the refinements grammar.
For example, \cref{fig:syntax-survey-answer-method} shows two possible syntaxes for refinements in methods.
The first one adds the refinements before each basic type (before the parameters and the return type), while the second only adds one annotation with all the refinements and is inspired by refinements of functional programming languages. The answers show a preference for the first syntax over the second since the first has a higher rate of "Preferable" answers and a lower rate of "Not Acceptable" answers. Similar analyses were applied to the remaining features leading to the presented in \cref{subsec:liquidjava_features}.results

\subsection{LiquidJava features} \label{subsec:liquidjava_features}
LiquidJava supports the refinement of variables (as shown in \cref{code:simple}), fields and methods (both parameters and return types). \Cref{code:method} shows an example of a method with the second parameter and return types extended with refinements.

\begin{lstlisting}[caption=Annotation of method \textit{inRange} with refinements and the verification of the method's invocations., label=code:method, basicstyle=\small,float=htb]
@Refinement("_ >= a && _ <= b")
public static int inRange(int a, @Refinement("b > a") int b){
    return a + 1;
}
...
inRange(10, 20); //correct
inRange(10, 2); /* Refinement Type Error 
                Type expected: (b > a); 
                Refinement found: (b == 2) && (a == 10)*/
\end{lstlisting}

Furthermore, refinements can be used in classes to model the state of their objects.
Classes are considered the fundamental programming elements of the Java language~\cite{java_book}. Although classes themselves do not have a specific value that can be refined, they can have methods that produce changes to the internal state of the objects. In this view, we can refine the object state when a method is called and when the method has ended with the annotation \lstinline|@StateRefinement(from="predicate", to = "predicate")|. 

\begin{figure}[htb]
\begin{minipage}{0.65\textwidth}
  \begin{lstlisting}[caption=Socket class object state refinement, label=code:socket, basicstyle=\small]
@ExternalRefinementsFor("java.net.Socket")
@StateSet({"unconnected", "bound", "connected", "closed"})
public interface SocketRefinements {
    @StateRefinement(to="unconnected(this)")
    public void Socket(); 

    @StateRefinement(from="unconnected(this)",to="bound(this)")
    public void bind(SocketAddress add);
    
    @StateRefinement(from="bound(this)", to="connected(this)")
    public void connect(SocketAddress add, int timeout);
    
    @StateRefinement(from="connected(this)")
    public void sendUrgentData(int n);
    
    @StateRefinement(to="closed(this)")
    public void close();
}
  \end{lstlisting}
\end{minipage}
\begin{minipage}{0.33\textwidth}
  \centering
  \includegraphics[scale=0.5]{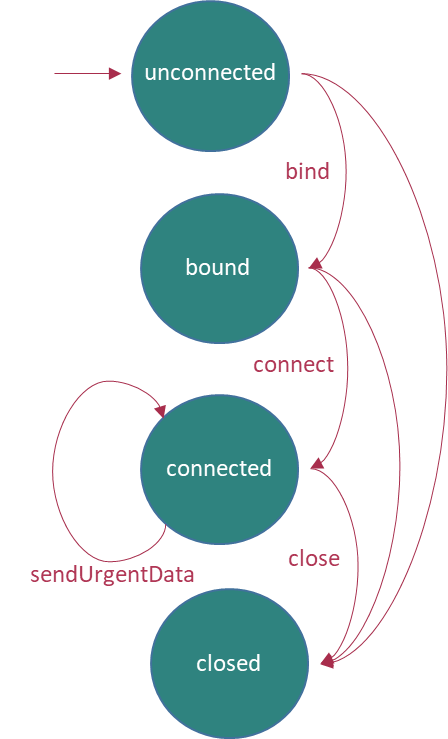} 
  \captionof{figure}{DFA that represents the Socket class states and transitions.}
  \label{fig:socket-dfa}
\end{minipage}
\end{figure}

\Cref{code:socket} presents the modeling of the class Socket from the external library java.net and whose methods follow an implicit order of invocations that the DFA of \cref{fig:socket-dfa} can describe. To model this class, we can start by defining the possible disjoint states with the annotation @StateSet and then, for each method, describe the allowed transitions. For example, the method \lstinline|sendUrgentData| can only be invoked if the object is connected, which means that this method should only be called after the \lstinline|connect| method. The developers using the Socket class can now detect if the invocations follow the correct protocol before executing the program. 

\subsection{Integration with IDE}\label{subsec:ide}
Integrated development environments (IDEs) are indispensable tools for software development nowadays. They integrate editor services that enhance developers productivity, such as content completion, documentation popups, and real-time type checking \cite{DBLP:conf/oopsla/KatsVKV12}. Overall, IDEs provide instant feedback to the developers while implementing the code, helping them correct errors before executing the programs.

To enhance the usability of LiquidJava, we have created an IDE plugin to allow the developers to use the lightweight verification integrated with the development environment.
To create the LiquidJava plugin, we used the Language Server Protocol~\cite{official_page_lsp} that decouples the creation of the language server from the implementation of the interface for the target editor. By using LSP we create a single implementation of the LiquidJava language server that can be paired with a client for any editor that implements LSP (e.g., Visual Studio Code~\cite{visual_studio_code_2016}, Eclipse~\cite{eclipse}, Emacs~\cite{emacs}). 

\begin{figure}[htb]
  \centering
  \includegraphics[scale=0.7]{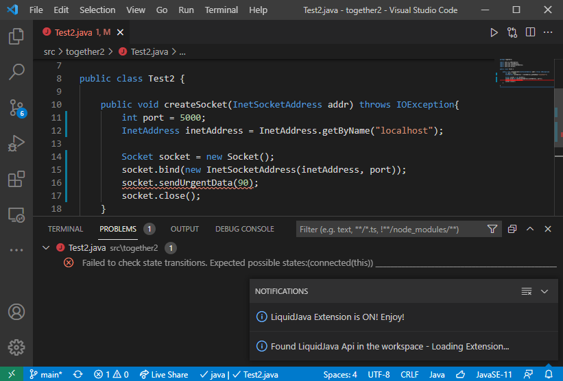}
  \caption{IDE Plugin reporting an error in the incorrect usage of the Socket class.}
  \label{fig:ide-plugin}
\end{figure}

For the implementation of the LiquidJava plugin, we chose to create a client for Visual Studio Code, given that it is one most used IDEs for Java development~\cite{ide_java_report_2021,ide_ecosystem_report_2021}. 
The main features of the plugin can be seen in \cref{fig:ide-plugin} and include:
\begin{itemize}
  \item \textbf{Error Reporting} -- This informs the user, in real-time, of the code elements whose specification could not be proved and underlines their exact location; 
  \item \textbf{Error Information} -- The problems tab shows the specification that LiquidJava failed to verify.
  For more detailed information, the user can hover the error and get all the verification conditions used to try to prove the specification. 
\end{itemize}

The plugin is available for download on the project website~\cite{liquidjava_website}, along with examples for the usage of refinements in variables, methods and classes.

\section{Evaluation}
To evaluate the usability of our approach, we conducted a user study with volunteer Java developers, to answer the following research questions:

\begin{enumerate}[label=\textbf{RQ. \arabic*}]
  \item \label{rq:intuitive} Are refinements easy to understand?
  \item \label{rq:errors} Is it easier and faster to find implementation errors using LiquidJava than with plain Java?
  \item \label{rq:annotation} How hard is it to annotate a program with refinements?
  \item \label{rq:would-use} Are developers open to using LiquidJava in their projects?
\end{enumerate}

The study had 30 participants familiar with Java,
of which 80\% described themselves as being only \enquote{Vaguely Familiar} or \enquote{Not Familiar} with refinement types,
and 90\% had no previous contact with LiquidJava.
Each participant had an online synchronous study session with an interviewer and followed the tasks described in \ref{subsec:setup}. The results of the study are presented in \ref{subsec:results}.

\subsection{Study Configuration} \label{subsec:setup}

We conducted the synchronous sessions through the Zoom video platform,
and gave the participants a survey with the study guidelines and answer placeholders,
and a GitHub repository with the study files.
The participants were also asked to share their screen with the VSCode editor and the document with the answers. 

The study was divided into six parts, as follows:
\begin{enumerate}
    \item \textbf{Task 1: Find the error in plain Java} --
    Participants had to find and fix semantic errors in two Java programs, where the implementation did not correspond to the informal documentation presented in the associated Javadoc.
    \item \textbf{Task 2: Understand Refinements without prior explanation} --
    Participants had to interpret the refinements present in different sections of the code (variables, methods and classes) and use them correctly and incorrectly.
    For example, the first exercise had a variable with a refinement type annotation and the participants had to assign both a correct and then an incorrect value.
    The results on this task aim to answer \ref{rq:intuitive} by counting the correct uses of the refinements. 
    \item \textbf{Overview of LiquidJava} --
    We introduced participants to LiquidJava using a 4-minute video and a webpage~\cite{liquidjava_website} explaining the examples of the previous task. Both resources were then available to be used within the remainder of the study.
    In the rest of the study, participants used LiquidJava through an IDE extension created for Visual Studio Code;
    \item \textbf{Task 3: Find the error with LiquidJava} --
    Similar to the Task 1, participants had to find and fix the incorrect behaviour of the programs.
    However, for this task, they were aided by the LiquidJava plugin.
    This task, paired with the Task 1, intend to answer \ref{rq:errors}. 
    The exercises were the same in both tasks, but they were split into two sets so that each participant could have different exercises in each task. Hence, half the participants had one set of exercises for Task 1 and a different one for Task 3, and the remaining half had the reverse set order. Therefore, the plain Java results serve as a baseline for the LiquidJava results. 
    \item \textbf{Task 4: Annotate Java programs with LiquidJava} --
    Participants were asked to add LiquidJava annotations to three Java programs that targeted the LiquidJava features of refinements on variables, fields, methods and classes. This part targets \ref{rq:annotation}, and includes a final question about the difficulty of adding the annotations. 
    \item \textbf{Final Comments} --
    Participants had the opportunity to express their thoughts on the overall experience of using LiquidJava by sharing what they most liked and disliked about the framework and whether they would use LiquidJava in their projects.
    This overview aimed to answer the last research question, \ref{rq:would-use}, and provide feedback to improve the project.
\end{enumerate}

\subsection{Results} \label{subsec:results}

The answers were reviewed and evaluated, leading to the results presented in this section.
We evaluated the answers to the presented problems
using the categories: \textit{Correct}, \textit{Incorrect}, \textit{Unanswered}, and \textit{Compiler Correct}.
\textit{Compiler correct} represents the answers that lead to the expected behaviour of the LiquidJava compiler without being completely correct according to the exercise.

\begin{figure*}[htb]
  \centering
  \begin{subfigure}[t]{0.5\textwidth}
      \centering
      \includegraphics[scale=0.35]{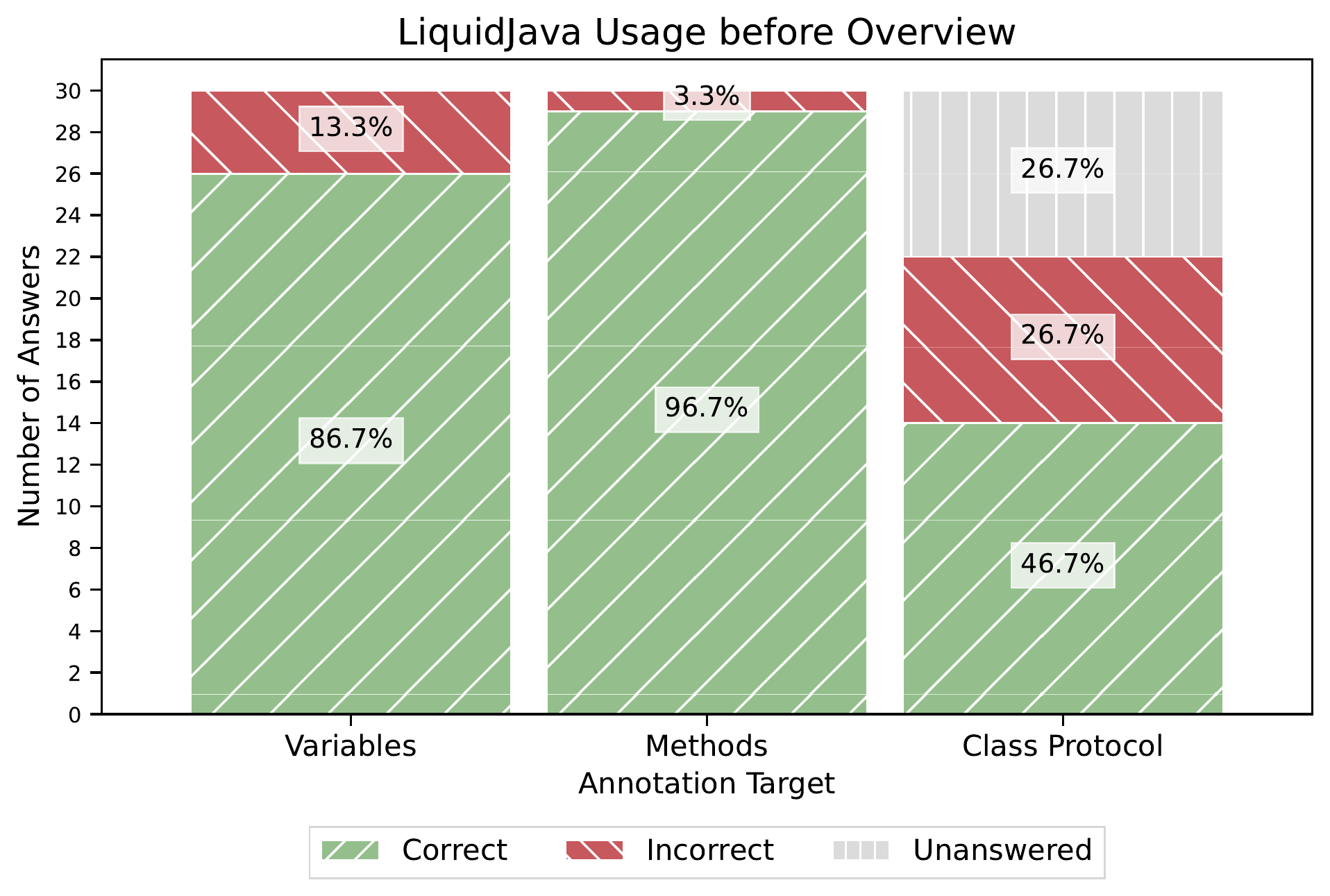}
      \caption{Answers to Understanding Refinements without prior explanation.}
      \label{fig:before_overview}
  \end{subfigure}%
  ~
  \begin{subfigure}[t]{0.57\textwidth}
      \centering
      \includegraphics[scale=0.35]{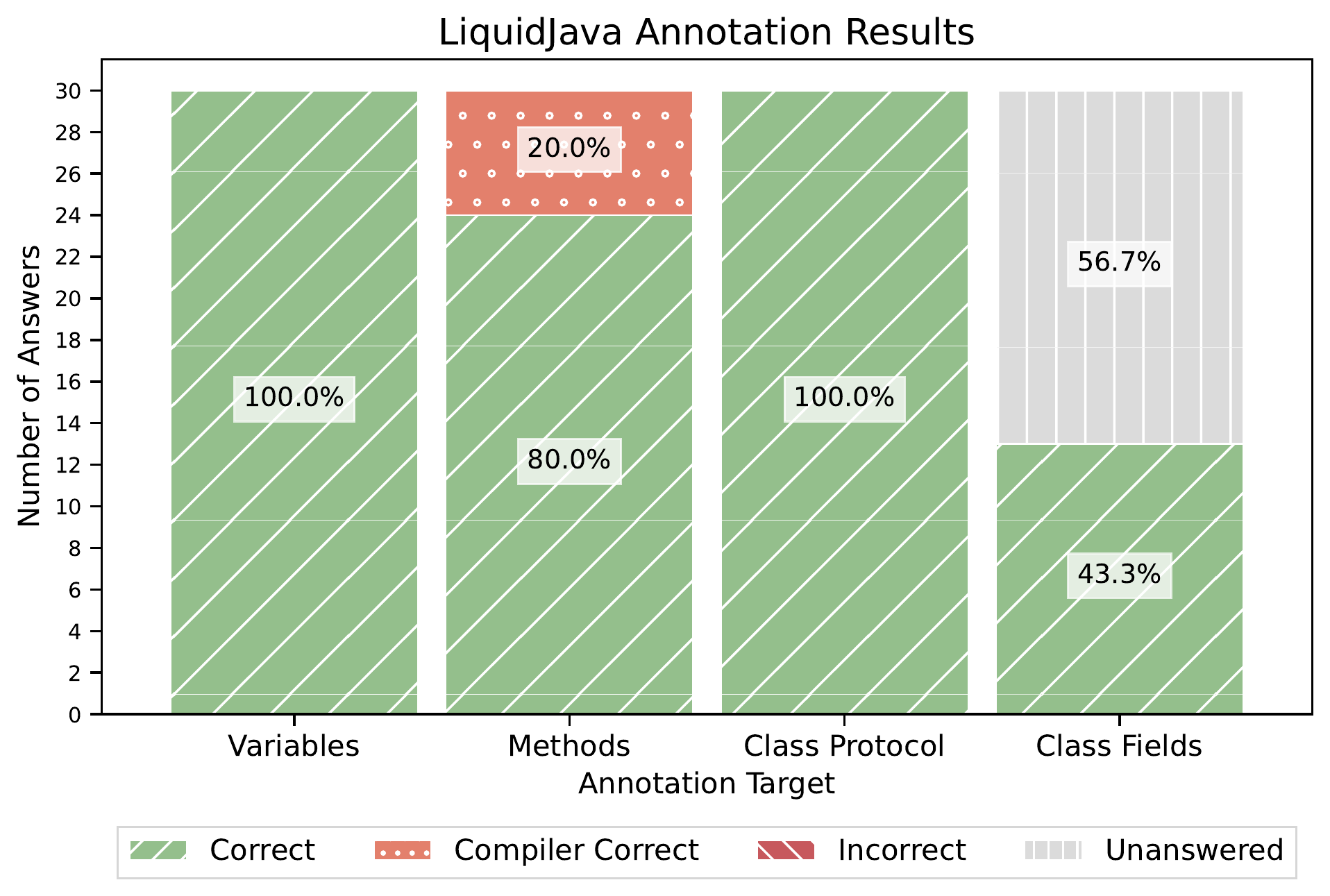}
      \caption{Answers to Annotation with LiquidJava.}
      \label{fig:annotate}
  \end{subfigure}
  \caption{Understandability of LiquidJava.}
  \label{fig:survey-understandability}
\end{figure*}

\Cref{fig:before_overview} shows that the refinements on variables and methods were intuitive to use since a large majority of participants answered correctly.
The refinements in the class were less intuitive, and 26.7\% of the participants even decided to leave this question unanswered.
However, after the short overview on LiquidJava, all participants could annotate the class with the protocol using LiquidJava (\cref{fig:annotate}), providing support for the notion that they are easy to learn and understand (\ref{rq:intuitive}).
When participants were asked to rate the ease of adding annotations on a scale of 0 (very difficult) to 5 (very easy),
60\% assigned it a value of 5, while the remaining 40\% chose the value 4.
This result provides evidence that it is fairly easy to add the refinement to the code (\ref{rq:annotation}).

\begin{figure*}[htb]
  \centering
  \begin{subfigure}[t]{0.5\textwidth}
      \centering
      \includegraphics[scale=0.35]{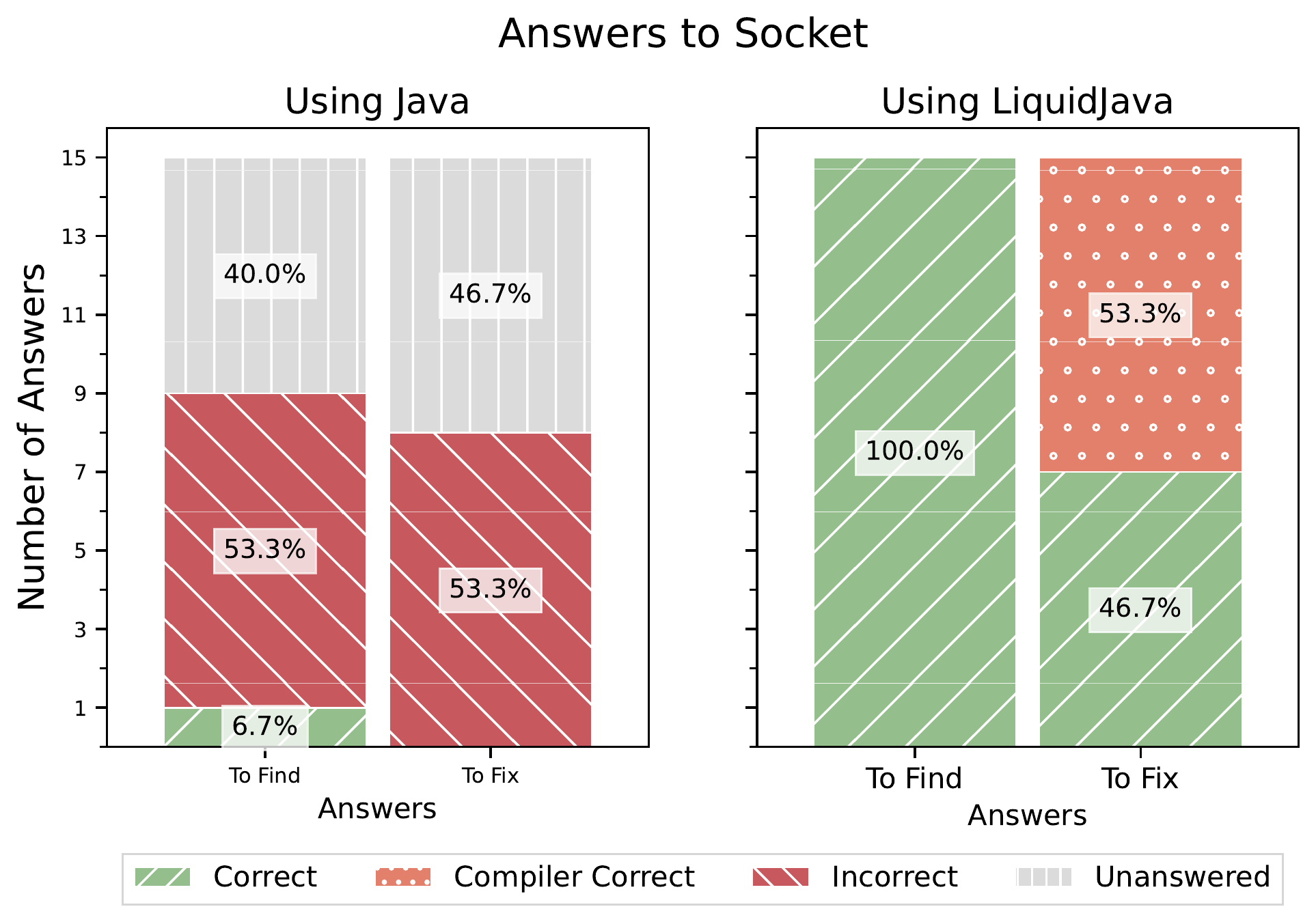}
      \caption{Answers on finding and fixing bug in Socket example.}
      \label{fig:results-socket}
  \end{subfigure}%
  ~
  \begin{subfigure}[t]{0.57\textwidth}
      \centering
      \includegraphics[scale=0.45]{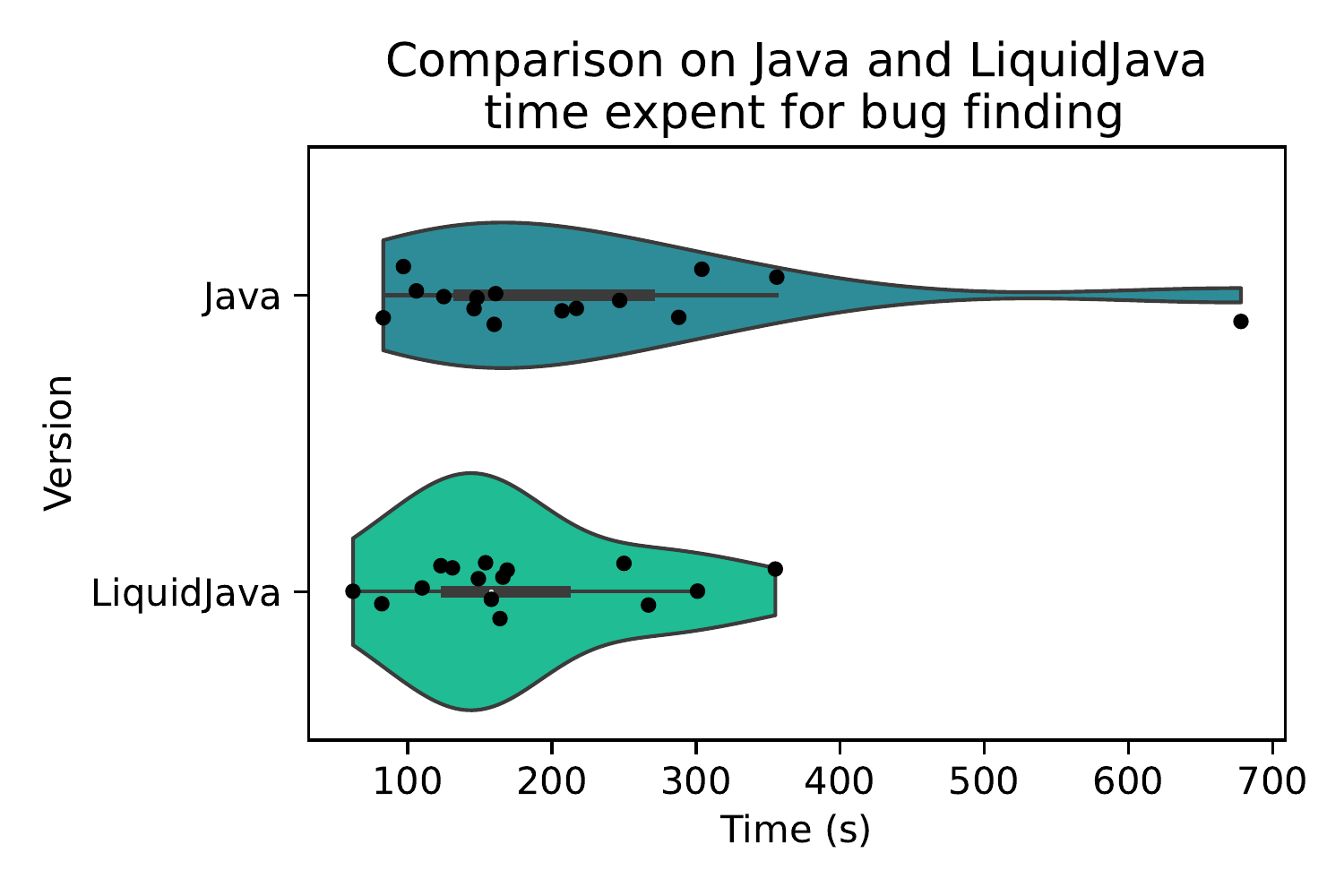}
      \caption{Time for finding and fixing bug in ArrayDeque example.}
      \label{fig:annotate}
  \end{subfigure}
  \caption{Using Java and LiquidJava to find and fix implementation bugs.}
  \label{fig:survey-time-arraydeque}
\end{figure*}

In Tasks 1 and 3, two programs were used.
The first one consisted of identifying the wrong order in which methods are called (\cref{code:socket}).
As shown in \cref{fig:results-socket}, almost no participant could find or fix the error without the help of LiquidJava.
In the second program, also with a wrong sequence of invocations of ArrayDeque methods, every participant was able find and fix the bugs in both Java and LiquidJava, but they were faster when using LiquidJava (\cref{fig:survey-time-arraydeque}). 
This result shows that LiquidJava might be more useful when added to lesser-known classes and protocols, and can reduce the time spent on the localization of the bugs, answering to \ref{rq:errors}.

On the overview of the experience, participants reported that they mostly enjoyed LiquidJava's error reporting, state refinement, and the syntax of its language.
For the features that the participants disliked, they referred to some aspects of the syntax and some plugin features, but the majority answered that there was nothing they disliked in the usage of LiquidJava.

The last question of the study was \textit{Would you use LiquidJava in your projects?} to which all the participants answered yes, which gives confidence that participants find this tool accessible for its gains (\ref{rq:would-use}).

\section{Conclusion}
This work presents the design of LiquidJava, a project focused on promoting the usability of refinement types, specifically in the Java language.
To this end, we defined design requirements and used the input of programmers familiar with Java to guide the syntax of the refinements language.
The features of the LiquidJava join the existing refinements concepts with the Java features, introducing the refinements of object state.
An implementation of LiquidJava was created and integrated within the Visual Studio Code editor.
We developed a research study to evaluate LiquidJava, and its results indicate that the refinements are easy to understand and that developers are interested in using LiquidJava in their projects. 

\bibliographystyle{ACM-Reference-Format}
\bibliography{references}


\begin{thebibliography}{23}


\ifx \showCODEN    \undefined \def \showCODEN     #1{\unskip}     \fi
\ifx \showDOI      \undefined \def \showDOI       #1{#1}\fi
\ifx \showISBNx    \undefined \def \showISBNx     #1{\unskip}     \fi
\ifx \showISBNxiii \undefined \def \showISBNxiii  #1{\unskip}     \fi
\ifx \showISSN     \undefined \def \showISSN      #1{\unskip}     \fi
\ifx \showLCCN     \undefined \def \showLCCN      #1{\unskip}     \fi
\ifx \shownote     \undefined \def \shownote      #1{#1}          \fi
\ifx \showarticletitle \undefined \def \showarticletitle #1{#1}   \fi
\ifx \showURL      \undefined \def \showURL       {\relax}        \fi
\providecommand\bibfield[2]{#2}
\providecommand\bibinfo[2]{#2}
\providecommand\natexlab[1]{#1}
\providecommand\showeprint[2][]{arXiv:#2}

\bibitem[\protect\citeauthoryear{??}{ema}{[n.d.]}]%
        {emacs}
 \bibinfo{year}{[n.d.]}\natexlab{}.
\newblock \bibinfo{title}{Gnu emacs}.
\newblock
\newblock
\urldef\tempurl%
\url{https://www.gnu.org/software/emacs/}
\showURL{%
\tempurl}


\bibitem[\protect\citeauthoryear{??}{off}{[n.d.]}]%
        {official_page_lsp}
 \bibinfo{year}{[n.d.]}\natexlab{}.
\newblock \bibinfo{title}{Language Server Protocol}.
\newblock
\newblock
\urldef\tempurl%
\url{https://microsoft.github.io/language-server-protocol/}
\showURL{%
\tempurl}


\bibitem[\protect\citeauthoryear{??}{vis}{2016}]%
        {visual_studio_code_2016}
 \bibinfo{year}{2016}\natexlab{}.
\newblock \bibinfo{title}{Visual Studio Code}.
\newblock
\newblock
\urldef\tempurl%
\url{https://code.visualstudio.com/}
\showURL{%
\tempurl}


\bibitem[\protect\citeauthoryear{??}{ide}{2021a}]%
        {ide_java_report_2021}
 \bibinfo{year}{2021}\natexlab{a}.
\newblock \bibinfo{title}{2021 Java technology report}.
\newblock
\newblock
\urldef\tempurl%
\url{https://www.jrebel.com/blog/2021-java-technology-report}
\showURL{%
\tempurl}


\bibitem[\protect\citeauthoryear{??}{ide}{2021b}]%
        {ide_ecosystem_report_2021}
 \bibinfo{year}{2021}\natexlab{b}.
\newblock \bibinfo{title}{JVM ecosystem REPORT 2021}.
\newblock
\newblock
\urldef\tempurl%
\url{https://snyk.io/jvm-ecosystem-report-2021/}
\showURL{%
\tempurl}


\bibitem[\protect\citeauthoryear{Bengtson, Bhargavan, Fournet, Gordon, and
  Maffeis}{Bengtson et~al\mbox{.}}{2008}]%
        {bengtson2008refinements}
\bibfield{author}{\bibinfo{person}{Jesper Bengtson},
  \bibinfo{person}{Karthikeyan Bhargavan}, \bibinfo{person}{C{\'{e}}dric
  Fournet}, \bibinfo{person}{Andrew~D. Gordon}, {and} \bibinfo{person}{Sergio
  Maffeis}.} \bibinfo{year}{2008}\natexlab{}.
\newblock \showarticletitle{Refinement Types for Secure Implementations}. In
  \bibinfo{booktitle}{\emph{Proceedings of the 21st {IEEE} Computer Security
  Foundations Symposium, {CSF} 2008}}. \bibinfo{publisher}{{IEEE} Computer
  Society}, \bibinfo{pages}{17--32}.
\newblock
\urldef\tempurl%
\url{https://doi.org/10.1109/CSF.2008.27}
\showDOI{\tempurl}


\bibitem[\protect\citeauthoryear{Burnay, Lopes, and Vasconcelos}{Burnay
  et~al\mbox{.}}{2020}]%
        {DBLP:conf/sefm/BurnayLV20}
\bibfield{author}{\bibinfo{person}{Nuno Burnay}, \bibinfo{person}{Ant{\'{o}}nia
  Lopes}, {and} \bibinfo{person}{Vasco~T. Vasconcelos}.}
  \bibinfo{year}{2020}\natexlab{}.
\newblock \showarticletitle{Statically Checking {REST} {API} Consumers}. In
  \bibinfo{booktitle}{\emph{Software Engineering and Formal Methods - 18th
  International Conference}}, \bibfield{editor}{\bibinfo{person}{Frank~S.
  de~Boer} {and} \bibinfo{person}{Antonio Cerone}} (Eds.),
  Vol.~\bibinfo{volume}{12310}. \bibinfo{pages}{265--283}.
\newblock
\urldef\tempurl%
\url{https://doi.org/10.1007/978-3-030-58768-0\_15}
\showDOI{\tempurl}


\bibitem[\protect\citeauthoryear{{Chugh R., Herman D., Jhala R.}}{{Chugh R.,
  Herman D., Jhala R.}}{[n.d.]}]%
        {DJS}
\bibfield{author}{\bibinfo{person}{{Chugh R., Herman D., Jhala R.}}}
  \bibinfo{year}{[n.d.]}\natexlab{}.
\newblock \bibinfo{title}{{Dependent Types for JavaScript}}.
\newblock
  \bibinfo{howpublished}{\url{http://goto.ucsd.edu/~ravi/research/oopsla12-djs.pdf}}.
\newblock


\bibitem[\protect\citeauthoryear{F{\"{a}}hndrich and Leino}{F{\"{a}}hndrich and
  Leino}{2003}]%
        {DBLP:conf/oopsla/FahndrichL03}
\bibfield{author}{\bibinfo{person}{Manuel F{\"{a}}hndrich} {and}
  \bibinfo{person}{K.~Rustan~M. Leino}.} \bibinfo{year}{2003}\natexlab{}.
\newblock \showarticletitle{Declaring and checking non-null types in an
  object-oriented language}. In \bibinfo{booktitle}{\emph{Proceedings of the
  2003 {ACM} {SIGPLAN} Conference on Object-Oriented Programming Systems,
  Languages and Applications, {OOPSLA} 2003}},
  \bibfield{editor}{\bibinfo{person}{Ron Crocker} {and} \bibinfo{person}{Guy
  L.~Steele Jr.}} (Eds.). \bibinfo{publisher}{{ACM}},
  \bibinfo{pages}{302--312}.
\newblock
\urldef\tempurl%
\url{https://doi.org/10.1145/949305.949332}
\showDOI{\tempurl}


\bibitem[\protect\citeauthoryear{Foundation}{Foundation}{[n.d.]}]%
        {eclipse}
\bibfield{author}{\bibinfo{person}{Eclipse Foundation}.}
  \bibinfo{year}{[n.d.]}\natexlab{}.
\newblock \bibinfo{title}{Eclipse IDE 2021-06: The Eclipse Foundation}.
\newblock
\newblock
\urldef\tempurl%
\url{https://www.eclipse.org/eclipseide/}
\showURL{%
\tempurl}


\bibitem[\protect\citeauthoryear{Freeman and Pfenning}{Freeman and
  Pfenning}{1991}]%
        {refinements-ml}
\bibfield{author}{\bibinfo{person}{Timothy~S. Freeman} {and}
  \bibinfo{person}{Frank Pfenning}.} \bibinfo{year}{1991}\natexlab{}.
\newblock \showarticletitle{Refinement Types for {ML}}. In
  \bibinfo{booktitle}{\emph{Proceedings of the {ACM} SIGPLAN'91 Conference on
  Programming Language Design and Implementation (PLDI)}},
  \bibfield{editor}{\bibinfo{person}{David~S. Wise}} (Ed.).
  \bibinfo{publisher}{{ACM}}, \bibinfo{pages}{268--277}.
\newblock
\urldef\tempurl%
\url{https://doi.org/10.1145/113445.113468}
\showDOI{\tempurl}


\bibitem[\protect\citeauthoryear{Gamboa}{Gamboa}{2021}]%
        {liquidjava_website}
\bibfield{author}{\bibinfo{person}{Catarina Gamboa}.}
  \bibinfo{year}{2021}\natexlab{}.
\newblock \bibinfo{title}{LiquidJava - Project Website}.
\newblock
\newblock
\urldef\tempurl%
\url{https://catarinagamboa.github.io/liquidjava.html}
\showURL{%
\tempurl}


\bibitem[\protect\citeauthoryear{Jhala and Vazou}{Jhala and Vazou}{2020}]%
        {jhala2020refinement}
\bibfield{author}{\bibinfo{person}{Ranjit Jhala} {and} \bibinfo{person}{Niki
  Vazou}.} \bibinfo{year}{2020}\natexlab{}.
\newblock \showarticletitle{Refinement Types: {A} Tutorial}.
\newblock \bibinfo{journal}{\emph{CoRR}}  \bibinfo{volume}{abs/2010.07763}
  (\bibinfo{year}{2020}).
\newblock
\showeprint[arxiv]{2010.07763}
\urldef\tempurl%
\url{https://arxiv.org/abs/2010.07763}
\showURL{%
\tempurl}


\bibitem[\protect\citeauthoryear{Kats, Vogelij, Kalleberg, and Visser}{Kats
  et~al\mbox{.}}{2012}]%
        {DBLP:conf/oopsla/KatsVKV12}
\bibfield{author}{\bibinfo{person}{Lennart C.~L. Kats},
  \bibinfo{person}{Richard~G. Vogelij}, \bibinfo{person}{Karl~Trygve
  Kalleberg}, {and} \bibinfo{person}{Eelco Visser}.}
  \bibinfo{year}{2012}\natexlab{}.
\newblock \showarticletitle{Software development environments on the web: a
  research agenda}. In \bibinfo{booktitle}{\emph{{ACM} Symposium on New Ideas
  in Programming and Reflections on Software, Onward! 2012, part of {SPLASH}
  '12, Tucson, AZ, USA, October 21-26, 2012}},
  \bibfield{editor}{\bibinfo{person}{Gary~T. Leavens} {and}
  \bibinfo{person}{Jonathan Edwards}} (Eds.). \bibinfo{publisher}{{ACM}},
  \bibinfo{pages}{99--116}.
\newblock
\urldef\tempurl%
\url{https://doi.org/10.1145/2384592.2384603}
\showDOI{\tempurl}


\bibitem[\protect\citeauthoryear{Ken~Arnold}{Ken~Arnold}{2005}]%
        {java_book}
\bibfield{author}{\bibinfo{person}{David~Holmes Ken~Arnold, James~Gosling}.}
  \bibinfo{year}{2005}\natexlab{}.
\newblock \bibinfo{booktitle}{\emph{THE Java™ Programming Language, Fourth
  Edition}}.
\newblock \bibinfo{publisher}{Addison Wesley Professional}.
\newblock
\showISBNx{0-321-34980-6}


\bibitem[\protect\citeauthoryear{P.}{P.}{2012}]%
        {roleVV}
\bibfield{author}{\bibinfo{person}{Upadhyay P.}}
  \bibinfo{year}{2012}\natexlab{}.
\newblock \showarticletitle{The Role of Verification and Validation in System
  DevelopmentLife Cycle}.
\newblock \bibinfo{journal}{\emph{IOSR Journal of Computer Engineering
  (IOSRJCE)}} \bibinfo{volume}{5}, \bibinfo{number}{1} (\bibinfo{year}{2012}),
  \bibinfo{pages}{17--20}.
\newblock


\bibitem[\protect\citeauthoryear{Pawlak, Monperrus, Petitprez, Noguera, and
  Seinturier}{Pawlak et~al\mbox{.}}{2016}]%
        {DBLP:journals/spe/PawlakMPNS16}
\bibfield{author}{\bibinfo{person}{Renaud Pawlak}, \bibinfo{person}{Martin
  Monperrus}, \bibinfo{person}{Nicolas Petitprez}, \bibinfo{person}{Carlos
  Noguera}, {and} \bibinfo{person}{Lionel Seinturier}.}
  \bibinfo{year}{2016}\natexlab{}.
\newblock \showarticletitle{{SPOON:} {A} library for implementing analyses and
  transformations of Java source code}.
\newblock \bibinfo{journal}{\emph{Softw. Pract. Exp.}} \bibinfo{volume}{46},
  \bibinfo{number}{9} (\bibinfo{year}{2016}), \bibinfo{pages}{1155--1179}.
\newblock
\urldef\tempurl%
\url{https://doi.org/10.1002/spe.2346}
\showDOI{\tempurl}


\bibitem[\protect\citeauthoryear{Rondon, Kawaguchi, and Jhala}{Rondon
  et~al\mbox{.}}{2008}]%
        {liquidtypes}
\bibfield{author}{\bibinfo{person}{Patrick~Maxim Rondon}, \bibinfo{person}{Ming
  Kawaguchi}, {and} \bibinfo{person}{Ranjit Jhala}.}
  \bibinfo{year}{2008}\natexlab{}.
\newblock \showarticletitle{Liquid types}. In
  \bibinfo{booktitle}{\emph{Proceedings of the {ACM} {SIGPLAN} 2008 Conference
  on Programming Language Design and Implementation}},
  \bibfield{editor}{\bibinfo{person}{Rajiv Gupta} {and}
  \bibinfo{person}{Saman~P. Amarasinghe}} (Eds.). \bibinfo{publisher}{{ACM}},
  \bibinfo{pages}{159--169}.
\newblock
\urldef\tempurl%
\url{https://doi.org/10.1145/1375581.1375602}
\showDOI{\tempurl}


\bibitem[\protect\citeauthoryear{{Rondon P., Bakst A., Kawaguchi M., Jhala
  R.,}}{{Rondon P., Bakst A., Kawaguchi M., Jhala R.,}}{[n.d.]}]%
        {csolve}
\bibfield{author}{\bibinfo{person}{{Rondon P., Bakst A., Kawaguchi M., Jhala
  R.,}}.} \bibinfo{year}{[n.d.]}\natexlab{}.
\newblock \bibinfo{title}{{CSolve: Verifying C With Liquid Types}}.
\newblock
  \bibinfo{howpublished}{\url{http://goto.ucsd.edu/~rjhala/papers/csolve_verifying_c_with_liquid_types.pdf}}.
\newblock


\bibitem[\protect\citeauthoryear{Sadowski, Aftandilian, Eagle, Miller{-}Cushon,
  and Jaspan}{Sadowski et~al\mbox{.}}{2018}]%
        {DBLP:journals/cacm/SadowskiAEMJ18}
\bibfield{author}{\bibinfo{person}{Caitlin Sadowski}, \bibinfo{person}{Edward
  Aftandilian}, \bibinfo{person}{Alex Eagle}, \bibinfo{person}{Liam
  Miller{-}Cushon}, {and} \bibinfo{person}{Ciera Jaspan}.}
  \bibinfo{year}{2018}\natexlab{}.
\newblock \showarticletitle{Lessons from building static analysis tools at
  Google}.
\newblock \bibinfo{journal}{\emph{Commun. {ACM}}} \bibinfo{volume}{61},
  \bibinfo{number}{4} (\bibinfo{year}{2018}), \bibinfo{pages}{58--66}.
\newblock
\urldef\tempurl%
\url{https://doi.org/10.1145/3188720}
\showDOI{\tempurl}


\bibitem[\protect\citeauthoryear{Vazou, Seidel, Jhala, Vytiniotis, and
  Peyton-Jones}{Vazou et~al\mbox{.}}{2014}]%
        {vazou2014refinement}
\bibfield{author}{\bibinfo{person}{Niki Vazou}, \bibinfo{person}{Eric~L
  Seidel}, \bibinfo{person}{Ranjit Jhala}, \bibinfo{person}{Dimitrios
  Vytiniotis}, {and} \bibinfo{person}{Simon Peyton-Jones}.}
  \bibinfo{year}{2014}\natexlab{}.
\newblock \showarticletitle{Refinement types for Haskell}. In
  \bibinfo{booktitle}{\emph{ACM SIGPLAN Notices}}, Vol.~\bibinfo{volume}{49}.
  ACM, \bibinfo{pages}{269--282}.
\newblock


\bibitem[\protect\citeauthoryear{Wright, Winters, and Manshreck}{Wright
  et~al\mbox{.}}{2020}]%
        {SEGoogle}
\bibfield{author}{\bibinfo{person}{Hyrum Wright},
  \bibinfo{person}{Titus~Delafayette Winters}, {and} \bibinfo{person}{Tom
  Manshreck}.} \bibinfo{year}{2020}\natexlab{}.
\newblock \bibinfo{booktitle}{\emph{Software Engineering at Google}}.
\newblock


\bibitem[\protect\citeauthoryear{Xi and Pfenning}{Xi and Pfenning}{1998}]%
        {xi1998arraybounds}
\bibfield{author}{\bibinfo{person}{Hongwei Xi} {and} \bibinfo{person}{Frank
  Pfenning}.} \bibinfo{year}{1998}\natexlab{}.
\newblock \showarticletitle{Eliminating Array Bound Checking Through Dependent
  Types}. In \bibinfo{booktitle}{\emph{Proceedings of the {ACM} {SIGPLAN} '98
  Conference on Programming Language Design and Implementation (PLDI)}}.
  \bibinfo{publisher}{{ACM}}, \bibinfo{pages}{249--257}.
\newblock
\urldef\tempurl%
\url{https://doi.org/10.1145/277650.277732}
\showDOI{\tempurl}


\end{thebibliography}

\end{document}